\documentclass[aps,prl,twocolumn,groupedaddress,preprintnumbers]{revtex4}

\usepackage{amsmath}

\begin{document}

\preprint{ BRX-TH664,
CALT 68-2906}

\title{PM$\ =\ $EM : \  {Partially Massless Duality Invariance  }}


\author{S.~Deser}
\email[]{deser@brandeis.edu}
\affiliation{Lauritsen Lab, Caltech, Pasadena CA 91125 and Physics Department, Brandeis University, Waltham, MA 02454, USA}

\author{A.~Waldron}
\email[]{wally@math.ucdavis.edu}
\affiliation{Department of Mathematics, University of California, Davis, CA 95616, USA}


\date{\today}

\begin{abstract}
In $d=4$ de Sitter space, novel conformally invariant photon-like theories consistently  couple  to  charged matter. 
We show that these higher spin, maximal depth, partially massless systems enjoy a  Maxwellian, ``electric-magnetic'' duality.

\end{abstract}


\maketitle

\section{Introduction}

The notion of duality invariance, $$\vec E \to \vec B\, ,\qquad \vec B \to -\vec E\, ,$$  is almost coeval with Maxwell's equations themselves, although a proof of its validity awaited over a century~\cite{D&T,D}. In addition to countless generalizations of ``duality'' in field and string theory, it has led to an enormous variety of more precise analogs, in particular to spin~2~\cite{H&T,D}, then to all free massless (integer or half-integer) spin~$s>0$ systems in flat space~\cite{D&S}~\footnote{
However, besides being inapplicable to massive systems, duality also ceases to be an invariance of the massless models' nonlinear, Yang--Mills~\cite{D&T} and general relativity~\cite{D&S1} extensions, an exception being quadratic, conformal (Weyl)  gravity~\cite{D&N1}. It seems equally unlikely that any nonlinear extension of PM, such as the putative one of massive gravity~\cite{deR}, will be dual invariant.}.

In de Sitter (dS), electromagnetic (EM) interactions can be mediated by generalized Maxwell systems~\cite{PMEM}. These are the maximal depth 
partially massless (PM) fields of~\cite{D&N,DW} which enjoy many characteristics of EM such as lightlike propagation~\cite{PMc}, gauge invariance~\cite{DW}, conformal invariance~\cite{DW2} and  stability~\cite{DW1}. Unlike EM, these PM models describe higher spin~$s$  propagating helicities $\pm s, \ldots , \pm1$~\cite{DW}.
Here we show that they also enjoy duality invariance, whence our title.

 
\section{$\mbox{d}$S Maxwell Duality} 

For Maxwell systems, duality in dS (or even generally curved)  backgrounds~\footnote{The invariance of Maxwell in both Poincare and dS spaces shows that the results of~\cite{BH} 
and~\cite{DS} relating duality-  and Poincare- invariances must be suitably interpreted: using the Schwinger-Dirac stress-tensor commutation relations that define the latter already assumes Poincar\'e; conversely, just because the 3+1 action (1) obeys these relations does not mean it is in flat space.}
is formally obvious in a covariant form notation, $F\to *F$: it just interchanges the Maxwell equations  with the Bianchi identity, 
$$
\delta F = 0 = d F\, ,
$$
where $\delta=\star\,  d\,  \star$. To establish the symmetry correctly, that is within the action principle in dS, we could simply appeal to the conformal invariance of the four dimensional Maxwell theory in order to  employ the flat space proof of~\cite{D&T,D}. For our purposes, an explicit dS proof is needed for our generalization to PM. In the particular dS coordinate frame
\begin{equation}\label{dS}ds^2=-dt^2 + \exp\big(2\sqrt{\Lambda/3}\, t\big)\,  d\vec x{}^{\, 2}\, ,
\end{equation}
the  first order Maxwell action just reduces to
\begin{equation}\label{SEA}
S[E,A;\Lambda]=\int d^4 x \Big[E \dot A - \frac 12 \, e^{-\sqrt{\Lambda/3}\, t}\,  \big\{E^2 + B(A)^2 \big\}\Big]\, .
\end{equation}
Here $E$ and $A$, are transverse by virtue of the Gau\ss\  constraint, {\it e.g.,} $A=A_i^T$, $\partial_i A^T_i=0$, and labels ``$\ {}^T_i\ $'' are henceforth dropped.
Also $B(A)$ denotes $\vec\nabla\times A$; for transverse~$V$ we have  $B(\vec \nabla \times V)=-\Delta V$.

Duality invariance is now exhibited as being a canonical transformation interchanging (with a helicity twist) the conjugate $(E,A)$ pair, while leaving $S(E,A;\Lambda)$ unchanged, but rotating $(E,B)$. Infinitesimally,
\begin{equation}\label{(2)}
\delta E = B(A)\, ,\quad \delta A = \vec\nabla\times  \big(\Delta^{\!-2} E\big) \Longrightarrow \delta B(A)= - E\, .       
\end{equation}
[Spatial non-locality is of course allowed.] Invariance of~(\ref{SEA}) under~(\ref{(2)}) is nearly manifest: The kinetic integrands' variations are easily seen to be total derivatives--of the form $ V \cdot \vec \nabla \times \dot{ V}$ 
(basically $F^*F$, in covariant language) up to possible but harmless Coulomb~$\sim |r-r'|$ factors, while the Hamiltonian's $\{E^2 + B^2\}$ is the very embodiment of rotation invariance.

\section{PM Systems}

PM systems originate from free mass~$m$ fields propagating in de Sitter (dS) backgrounds ($\Lambda>0$)
for which special~$m\, \colon \Lambda$ tunings yield additional gauge invariance(s)~\cite{D&N,DW}, thereby eliminating one or more lower helicity components from the, unavoidable in flat space, $(2s+1)$~total. For maximal depth PM systems, the helicity zero excitation is thereby removed, leaving only helicity~$\pm(s,\ldots,1)$ modes. We simply write the final form of their  (gauge-invariant) actions when all constraints are solved. [This is the critical step that requires sourceless fields: in the original Maxwell example,~$\vec B$ is identically transverse ($\vec\nabla\cdot\vec B=0$), so duality rotation is only well-defined (let alone an invariance) when the electric field is likewise transverse, with vanishing longitudinal--Coulomb component. This is also why zero mass is required in flat space.] Reduced PM actions in terms of transverse-traceless (TT) tensors were first given in Eq. (29) of~\cite{DW1} for PM spin~2 or Eq. (24) of~\cite{DW2} for arbitrary~$s$, maximal depth PM fields:
\begin{widetext}
\begin{equation}\label{(1)}
S= \sum_{\varepsilon=1}^s S[\pi_{i_1\ldots i_\varepsilon}^{TT}, \varphi_{i_1\ldots i_\varepsilon}^{TT};\Lambda]\, ,\ \mbox{ where }\:\: 
S[p,q;\Lambda] := \int d^4x \Big[p\,  \dot q-\frac12\, \big\{ p^2 + e^{-2\sqrt{\Lambda/3}\, t}\, B(q)^2- \frac{\Lambda}{12}\,  q^2\big\}\Big]\, .
\end{equation}
\end{widetext}
Here the sum over $\varepsilon$ runs over the helicities $1,\ldots, s$ of a (maximal depth) partially massless field and thus avoids the dangerous helicity zero mode. All indices are suitably contracted in each helicity's action and $B(q):=\vec\nabla\times q$ denotes the ``magnetic'' field, namely   the symmetrized curl~\cite{DD,D&S} 
\begin{equation}\label{curl}\vec \nabla\times \varphi^{TT}_{i_1\ldots i_\varepsilon}:= \epsilon^{\phantom T\!\!}_{(i_1|kl}\partial^{\phantom T\!\!}_k \varphi_{l|i_2\ldots i_\varepsilon)}^{TT}\, .
\end{equation}  
Note that (only) in 
dimension $d=3+1$  do the tensor ranks of each $B(\varphi)$ still match those of their potentials, and so of their corresponding  ``electric'' companions~$\pi$. In what follows, the (easily verified) identity for transverse-traceless tensors
$$
B(\vec\nabla\times q)=-\Delta q\, ,
$$
will play an essential {\it r\^ole}.

In the dS coordinates~(\ref{dS}) 
 used in~\cite{DW1,DW2}, the only metric dependence of the action~(\ref{(1)}) is through~$\Lambda$. Note also, as shown in~\cite{DW1}, although the Hamiltonian in~(\ref{(1)}) is neither 
time independent, nor manifestly positive, the generator of time translations, constructed from the composition $\xi^\mu T_{\mu\nu}$ of the timelike, dS Killing vector~$\xi^\mu$ and the stress energy tensor~$T_{\mu\nu}$~\cite{AD}, is both conserved and positive within the intrinsic horizon.

\section{PM Duality}

We now generalize the above scheme to PM. The essential point is that the duality rotations occur separately  within each helicity sector.
The key maneuver, therefore, is to bring the action $S[p,q;\Lambda]$ displayed in~(\ref{(1)}) to the manifestly duality invariant form
$S[E,A;\Lambda]$ of~(\ref{SEA}). This is achieved  via the field redefinition
$$
E:=e^{\frac12\sqrt{\Lambda/3}\, t}\, \Big\{p \ -\  \frac{\scriptstyle \sqrt{\Lambda/3}}2\,   q\Big\} \, ,\quad
A:=e^{-\frac12\sqrt{\Lambda/3}\, t}\, q\, .
$$
The proof  that duality invariance is a canonical transformation is now identical to that of the dS Maxwell theory given above, save that the vector curl is replaced by its higher rank symmetrized counterpart in~(\ref{curl}). PM's duality rotation invariance is, like Maxwell's, traceable to conformal invariance.

\section{Summary}
We have explicitly established the (extended) duality invariance of  
maximal depth PM systems.
Noting that our models live in dS, one might speculate on their possible cosmological relevance: to the extent that such PM fields might be present, their radiative interactions could have consequences in the corresponding era. However, their coupling to charges are sufficiently unusual~\cite{PMEM} that we have not yet tried to investigate this topic.

\begin{acknowledgments}
 S.D. was supported in part by NSF PHY- 1064302 and DOE DE-FG02-164 92ER40701 grants. 
\end{acknowledgments}


\end{document}